\documentclass[12pt]{article}
\begin{document}
\title{{\bf Why the Kirnberger Kernel Is So Small}
\thanks{Alberta-Thy-14-09, arXiv:0907.5249}}
\author{
Don N. Page
\thanks{Internet address:
don@phys.ualberta.ca}
\\
Theoretical Physics Institute\\
Department of Physics, University of Alberta\\
Room 238 CEB, 11322 -- 89 Avenue\\
Edmonton, Alberta, Canada T6G 2G7
}
\date{(2009 July 29)}

\maketitle
\large
\begin{abstract}
\baselineskip 18 pt

Defining the musical interval of the Kirnberger kernel, or Kirn-kern, to
be one-twelfth the atom of Kirnberger, or the difference between a grad
and a schisma, its natural logarithm, $k =
(161/12)\ln{2}-7\ln{3}-\ln{5}$, is extremely small, $k \approx
0.000\,000\,739\,401$. Here an explanation of this coincidence is given
by showing that $k = (1/6)(11\tanh^{-1}[(3/23)/11] -
21\tanh^{-1}[(3/23)/21]) \approx (2^5 5)/(3\cdot 7^2 11^2 23^3) \approx
0.000\,000\,739\,322$.

\end{abstract}
\normalsize

\baselineskip 13.9 pt

\newpage

\section*{Introduction}

The historical accident that the most common musical scale today divides
the octave into 12 semitones is no doubt due to the fact that 3 is close
to $2^{19/12}$, so that a justly tuned perfect fifth (pitch ratio 3:2 or
3/2) is very close to seven semitones of equal temperament (pitch ratio
$2^{7/12}$).  The difference has been defined by Andreas Werckmeister
(1645--1706) to be the {\it grad} \cite{schisma}, which (on a
logarithmic scale) is one-twelfth of the Pythagorean or ditonic comma
that is the difference between twelve perfect fifths and seven octaves
(pitch ratio $(3/2)^{12}/2^{7} = 3^{12}/2^{19} = 531\,441/524\,288
\approx 1.013\,643$).

Having divided up the octave into twelve semitones so that a pitch ratio
of three may be closely approximated by nineteen semitones, it is a
rather fortuitous ``accident'' that a pitch ratio of five may also be
rather closely approximated by an integer number of semitones, in this
case twenty-eight.  This is equivalent to the fact that a justly tuned
major third (pitch ratio $5/4 = 1.25$) is close to one-third of an
octave or four semitones of equal temperament (pitch ratio $2^{4/12} =
2^{1/3} \approx 1.259\,921$), that is, the fact that the cube root of
two is close to 5/4.  Leaving aside the part of the coincidence that
dividing up the octave into twelve semitones to get a good approximation
in equal temperament to the justly tuned perfect fifth gives the cube
root of two as an integer number of semitones, i.e., that the 3 denoting
the cube root is an integer divisor of 12, the further coincidence that
$2^{1/3} \sim 5/4$ or $2^7 \sim 5^3$ is the basis of the fact that a
kilobyte, $2^{10} = 1\,024$ bytes, is close to one thousand bytes.  That
is, the ratio of a kilobyte to one thousand bytes is $2^7/5^3 = 1.024$,
which in music is the pitch ratio called a diesis or minor diesis or
enharmonic diesis \cite{diesis}.

However, there is an interesting further coincidence with these ratios. 
This is the fact that the difference between four semitones of equal
temperament and a justly tuned major third (one third of a diesis on a
logarithmic scale) is just very slightly more than an integer number,
seven, times the difference between a justly tuned perfect fifth and
seven semitones of equal temperament (a grad).  The ratio between these
two differences on a logarithmic scale is
$\ln{(2^{7/3}/5)}/\ln{(3/2^{19/12})} \approx 7.000\,655$.  That is, on a
logarithmic scale, one-third of an enharmonic diesis is just very
slightly more than seven times a grad (which is one-twelfth of a
Pythagorean comma).  A consequence of this coincidence is that
$2^{23/12}/5^{1/7} \approx 3.000\,000\,316\,886$ is very nearly three. 
One can then easily see that another consequence is that the twelfth
root of two multiplied by the seventh root of five, which numerically to
twelve-decimal-place accuracy is $1.333\,333\,192\,495$, is within about
one part in ten million of being four-thirds.

This coincidence was known to Johann Kirnberger (1721--1783), a student
of Johann Sebastian Bach, who developed \cite{schisma} a rational
intonation approximation to equal temperament by flattening the justly
tuned perfect fifth by what is now called a {\it schisma}, the pitch
ratio $3^8 5/2^{15} = 32805/32768 = 1.001\,129\,150\,390\,625$, rather
than by the grad that is the pitch ratio $3/2^{19/12} \approx
1.001\,129\,890\,627\,526$ that would be needed to get give exact equal
temperament.  (The musical sense of the word {\it schisma} was
introduced by Alexander John Ellis (1814--1890), whom George Bernard
Shaw acknowledged as the prototype of Professor Henry Higgins in his
1912 play {\it Pygmalion} \cite{Ellis}.)  Therefore, the Kirnberger
fifth would be the ratio $2^{14}/(3^7 5) = 16384/10935 \approx
1.498\,308\,184\,728$ rather than the equal tempered fifth of $2^{7/12}
\approx 1.498\,307\,076\,877$.  Twelve of these Kirnberger fifths
exceeds seven octaves, and therefore fails to close exactly, by the tiny
interval of $2^{161}3^{-84}5^{-12} \approx 1.000\,008\,872\,860$, the
atom of Kirnberger.

\section{The size of the Kirnberger kernel}

Here I shall show why the atom of Kirnberger is so small, reducing the
coincidence noted above to an coincidence involving only integers.  For
simplicity, I shall define the {\it Kirnberger kernel}, or {\it
Kirn-kern} for short, to be (on a logarithmic scale) one-twelfth of the
atom of Kirnberger.  (The word {\it kernel} is used as the English
cognate of the German word {\it Kern}, meaning ``nucleus,'' that I am
here taking to be something smaller than an ``atom.'')

Now let me define lower-case letters for the natural logarithms of the
various musical intervals involved, and use the corresponding upper-case
letters for the values of these quantities in the musical unit of a cent
\cite{cent}, which was also introduced by the Henry Higgins prototype
Alexander Ellis and which corresponds to the musical interval of
one-hundredth of a semitone, $2^{1/1200} \approx 1.000\,577\,790$ and
whose natural logarithm is 
\begin{equation}
c \equiv \ln{(2)}/1200 \approx 0.000\,577\,622\,650.
\label{cent}
\end{equation}
First, define the natural logarithm for an equal-temperament semitone as
\begin{equation}
h \equiv \frac{1}{12}\ln{2} \approx 0.057\,762\,265\,047,
\label{semitone}
\end{equation}
with its value in cents being
\begin{equation}
H \equiv \frac{h}{c} = 100.
\label{semitone-cents}
\end{equation}
Second, define the natural logarithm of the grad as
\begin{equation}
g \equiv \ln{3}-\frac{19}{12}\ln{2} \approx 0.001\,129\,252\,782,
\label{grad}
\end{equation}
with its value in cents being
\begin{equation}
G \equiv \frac{g}{c} = 1.955\,000\,865\,387.
\label{grad-cents}
\end{equation}
Third, define the natural logarithm of the Kirnberger kernel as
\begin{equation}
k \equiv \frac{161}{12}\ln{2}-7\ln{3}-\ln{5}
 \approx 0.000\,000\,739\,402,
\label{Kirn-kern}
\end{equation}
with its value in cents being
\begin{equation}
K \equiv \frac{k}{c} = 0.001\,280\,077\,453.
\label{Kirn-cents}
\end{equation}
I shall refer to these quantities as the semitone $h$, the grad $g$, and
the Kirnberger kernel (or Kirn-kern) $k$, using the values of the
natural logarithms instead of the pitch ratios themselves.

The semitone $h$, grad $g$, and Kirn-kern $k$ give a basis for writing
the logarithm of any musical ratio formed from integer numbers of
equal-temperament semitones, just perfect fifths, and just major thirds
as a sum with integer coefficients of the $h$, $g$, and $k$.  In
particular,
\begin{eqnarray}
\ln{2} &=& 12h, \nonumber \\
\ln{3} &=& 19h + g, \nonumber \\
\ln{5} &=& 28h -7g -k,
\label{logs}
\end{eqnarray}
so
\begin{equation}
\ln{(2^{l/12} 3^m 5^n)} = (l+19m+28n)h + (m-7n)g -nk.
\label{expressed}
\end{equation}

These basis numbers are also of greatly differing sizes, $h \approx
51.15 g \approx 78120 k$ and $g \approx 1527 k$.  It is not very
surprising that the grad $g$ is much smaller than the semitone, since
the number of semitones in an octave was chosen to make it small, but it
is rather surprising that the Kirn-kern is so much smaller even than the
grad and that the ratio of the Kirn-kern to the grad is even smaller (by
a factor of nearly thirty) than the ratio of the grad to the semitone. 
It is this smallness that I wish to explain in terms of a somewhat less
surprising exact coincidence between integers.

Define the natural logarithm of the enharmonic diesis (pitch ratio
128/125) to be
\begin{equation}
d \equiv \ln{\frac{128}{125}} = 7\ln{2}-3\ln{5} = 21g + 3k,
\label{diesis}
\end{equation}
and the natural logarithm of the syntonic comma (pitch ratio 81/80) to
be
\begin{equation}
s \equiv \ln{\frac{81}{80}} = -4\ln{2}+4\ln{3}-\ln{5} = 11g + k.
\label{syntonic}
\end{equation}
Since these musical intervals are considerably smaller than a semitone
(being 41.06 and 21.51 cents respectively), it is not surprising that
they can be written entirely in terms of the grad $g$ and Kirn-kern $k$.

One can readily solve these two equations to get the Kirn-kern $k$ in
terms of the enharmonic diesis $d$ and syntonic comma $s$:
\begin{equation}
k = \frac{11}{12}d-\frac{7}{4}s
  = \frac{11}{12}\ln{\frac{128}{125}}-\frac{7}{4}\ln{\frac{81}{80}}.
\label{Kirn}
\end{equation}

Next, we use the fact that the logarithm of a number just a bit larger
than unity can be written in terms of the inverse hyperbolic tangent of
a corresponding small number, which one can write as a power series in
that small number.  In particular, if we write $128/125 = (1+x)/(1-x)$
and $81/80 = (1+y)/(1-y)$, then $x = (128-125)/(128+125) = 3/253 =
3/(11\cdot 23)$ and $y = (81-80)/(81+80) = 1/161 = 1/(7\cdot 23) =
3/(21\cdot 23)$, and one may write
\begin{equation}
d = \ln{\frac{1+x}{1-x}} = 2\tanh^{-1}{x}
 = 2\left(x + \frac{1}{3}x^3 + \cdots \right),
\label{dies}
\end{equation}
\begin{equation}
s = \ln{\frac{1+y}{1-y}} = 2\tanh^{-1}{y}
 = 2\left(y + \frac{1}{3}y^3 + \cdots \right).
\label{synton}
\end{equation}
Then one gets
\begin{eqnarray}
k &=& \frac{1}{6}\left(11\tanh^{-1}{\frac{3}{11\cdot 23}}
           -21\tanh^{-1}{\frac{3}{21\cdot 23}}\right) \nonumber \\
  &=& \frac{1}{6}\left(11\frac{3}{11\cdot 23}-21\frac{3}{21\cdot 23}
  +\frac{1}{3}\left(11\left(\frac{3}{11\cdot 23}\right)^3
                   -21\left(\frac{3}{21\cdot 23}\right)^3\right)
		   +\cdots\right) \nonumber \\
  &\approx& \frac{2^5 5}{3\cdot 7^2 11^2 23^3} = \frac{160}{216414429}
            \nonumber \\
  &\approx& 0.000\,000\,739\,322,
\label{Kir}
\end{eqnarray}
which is very tiny and is also within one part in nine thousand of the
correct numerical value for the Kirnberger kernel, $k \approx
0.000\,000\,739\,402$.

That is, the Kirnberger kernel or Kirn-kern $k$ (the natural logarithmic
difference between a grad and a schisma, one-twelfth the atom of
Kirnberger) is very small because $11x=21y$, so that the first-order
terms in the expansion of the inverse hyperbolic tangents cancel, and
one is left only with cubic and higher terms in the small quantities $x
= 3/253$ and $y = 1/161$.  Thus the coincidence reduces to the fact that
$11(2^7-5^3)/(2^7+5^3)=21(3^4-2^4 5)/(3^4+2^4 5)$, or to the fact that
the integers $11(2^7-5^3)(3^4+2^4 5)$ and $21(3^4-2^4 5)(2^7+5^3)$
(both being 5313) are equal.  Alternatively, if we do not start with
knowing the particular integer (7) that is the approximation to the
ratio $R = \ln{(2^{7/3}/5)}/\ln{(3/2^{19/12})}$ but just use the fact
that $R = 4d/(3s-d)$ and want to show that it is close to being an
integer by using the first term in the expansion of the inverse
hyperbolic tangent expression for the enharmonic diesis $d$ and for the
syntonic comma $s$, then if we use the facts that $3^4-2^4 5 = 81-80 =
1$ and that $2^7 - 5^3 = 128-125 = 3$, the coincidence reduces to the
fact that the approximation for $R$ becomes $(2^5 5 + 1)/(2^6-2^3 5 - 1)
= 161/23 = 7$, which is an integer.

Now it is still not obvious whether or not there is a simple reason I
could see in my head why $(2^5 5 + 1)/(2^6-2^3 5 - 1)$ is an integer,
but it is easy to check mentally, whereas for me it would be a lot more
work to check directly, even using pen and paper, without using a
calculator or computer, that $R = \ln{(2^{7/3}/5)}/\ln{(3/2^{19/12})}$
is very close to an integer.  Therefore, why one gets the ratio $161/23$
is still not very transparent to me, but at least I can easily see that
it is an integer and that it is perhaps not too surprising that a
rational number with a denominator as small as 23 is {\it a priori} not
too unlikely to be an integer.  (If the numerator were random with a
uniform measure in some large range of integers, the chance would be 1
in 23, or a bit more than 4\%.)  Therefore, the present analysis does
remove some of the mystery in my own mind as to why the Kirnberger
kernel is so small.

\section{Rational approximations for the grad, semitone, and other
logarithms}

Having found that one may easily define the rational approximation $k
\approx (2^5 5)/(3\cdot 7^2 11^2 23^3)$ for the Kirnberger kernel or
Kirn-kern $k$, one might also be interested in deriving similar simple
rational approximations for the grad $g$ and semitone $h$.  One can see
that the grad is $g = s/4 - d/12$, and in this case the first order
terms in the expansion of the inverse hyperbolic tangent expressions for
$s$ and $d$ do not cancel.  If one uses just these first-order
expressions, one gets
\begin{equation}
g \approx \frac{2}{7\cdot 11\cdot 23} = \frac{2}{1771}
 \approx 0.001\,129\,305\,477,
\label{gra}
\end{equation}
which is within one part in twenty-one thousand of the correct numerical
value of $g \approx 0.001\,129\,252\,782$.
From this approximation for the logarithmic interval $g$ of the grad,
one can also calculate an approximation to second order in $g$ for the
pitch ratio, $e^g \approx (1+g/2)/(1-g/2) \approx (1771+1)/(1771-1) =
886/885$.

To get an analogous rational estimate for the semitone, one can take the
first term in the expansion for the logarithm of the classic chromatic
semitone or minor chroma \cite{intervals},
\begin{equation}
t \equiv \ln{\frac{25}{24}} \approx \frac{2}{49}.
\label{chroma}
\end{equation}
Then combining this with the approximation above for the grad $g$ and
neglecting the tiny contribution from the Kirn-kern $k$ gives
\begin{equation}
h = t + 15g + 2k \approx \frac{2}{7^2}
 + \frac{2\cdot 3\cdot 5}{7\cdot 11\cdot 23}
 = \frac{716}{7^2 11\cdot 23} = \frac{716}{12397}
 \approx 0.057\,755\,908\,688,
\label{semi}
\end{equation}
which is within one part in nine thousand of the correct numerical value
of $h \approx 0.057\,762\,265\,047$.

From these rational approximations, one can readily get rational
approximations for natural logarithm of any product of integer powers of
an equal-temperament semitone ($2^{1/12}$), 3, and 5.  In particular,
\begin{eqnarray}
\ln{2} &=& 12h \approx \frac{8592}{12397}, \nonumber \\
\ln{3} &=& 19h + g \approx \frac{13618}{12397}, \nonumber \\
\ln{5} &=& 28h -7g -k \approx 28h-7g \approx \frac{19950}{12397},
\label{log-app}
\end{eqnarray}
so
\begin{equation}
\ln{(2^{l/12} 3^m 5^n)} \approx \frac{716l+13618m+19950n}{12397},
\label{expressed2}
\end{equation}
at least so long as one does not have both $12l+19m+28n = 0$ and $m-7n =
0$ so that the expression is simply equal to some multiple of $k$ that
is being neglected in this approximation.  (If one does have both
$12l+19m+28n = 0$ and $m-7n = 0$, then $\ln{(2^{l/12} 3^m 5^n)} \approx
-160n/216414429$ from the approximation above for $k$.)

If one instead wants approximations in terms of cents, one can use the
approximate value
\begin{equation}
c \equiv \ln{(2)}/1200 \equiv \frac{h}{100}
  \approx \frac{179}{5^2 7^2 11\cdot 23}
  \approx 0.000\,577\,559\,087
\label{cent-app}
\end{equation}
to get the rational approximations
\begin{eqnarray}
\ln{2} &=& 1200\ \mathrm{cents}, \nonumber \\
\ln{3} &=& 100\frac{\ln{3}}{\ln{2}} \approx \frac{340450}{179}
   \ \mathrm{cents}, \nonumber \\
\ln{5} &=& 100\frac{\ln{5}}{\ln{2}} \approx \frac{498750}{179}
   \ \mathrm{cents},
\label{log-app-cents}
\end{eqnarray}
so
\begin{equation}
\ln{(2^{l/12} 3^m 5^n)} \approx \frac{50}{179}(358l+6809m+9975n)
    \ \mathrm{cents}.
\label{expressed3}
\end{equation}

Alternatively, one can get rational approximations for the common
logarithms (logarithms to base ten) by dividing the rational
approximations for the natural logarithms by the rational approximation
for $\ln{10} = \ln{2} + \ln{5} = 40h -7g -k \approx 40h-7g \approx 2\cdot
3\cdot 67\cdot 71/7^2 11\cdot 23 = 28542/12397 \approx 2.302\,331$,
about one part in nine thousand smaller than the actual value $\ln{10}
\approx 2.302\,585$.  Then one gets
\begin{equation}
\log_{10}{(2^{l/12} 3^m 5^n)} \approx \frac{358l+6809m+9975n}{14271}.
\label{expressed10}
\end{equation}

One can now use this approximation to get rational approximations, with
the same denominator, for the common logarithms of all integers up
through ten, except for seven.  For seven, one may use the musical
interval of the Breedsma \cite{Breedsma}, with pitch interval 2401/2400,
the difference (on a logarithmic scale) between the septimal diesis
(pitch ratio 49/48) and the septimal sixth-tone (50/49), to motivate the
approximation $4\ln{7} = \ln{2401} = \ln{(2401/2400)} + \ln{2400}
\approx 2/4801 + 5\ln{2} + 3\ln{3} + 2\ln{5} \approx 2/4801 + 135h - 13g
\approx (5.2 + 96660 - 182)/12397 = 96483.2/12397$, or $\ln{7} \approx
24120.8/12397$.  Alternatively, one may use the ragisma \cite{ragisma},
with pitch ratio 4375/4374, the difference between the septimal minor
third (7/6) and two Bohlen-Pierce small semitones (27/25), to motivate
the approximation $\ln{7} = \ln{(4375/4374)} + \ln{2} + 7\ln{3} -
4\ln{5} \approx 2/8749 + 33h + 35g \approx (2.8 + 23628 + 490)/12397 =
24120.8/12397$.  Then one gets

\newpage

\begin{eqnarray}
\log_{10}{1} &=& \frac{0}{14271} = 0, \nonumber \\
\log_{10}{2} &\approx& \frac{4296}{14271}\approx 0.301\,030,\nonumber \\
\log_{10}{3} &\approx& \frac{6809}{14271}\approx 0.477\,121,\nonumber \\
\log_{10}{4} &\approx& \frac{8592}{14271}\approx 0.602\,060,\nonumber \\
\log_{10}{5} &\approx& \frac{9975}{14271}\approx 0.698\,970,\nonumber \\
\log_{10}{6} &\approx& \frac{11105}{14271}\approx 0.778\,151,\nonumber\\
\log_{10}{7} &\approx& \frac{12060.4}{14271}\approx 0.845\,098,\nonumber \\
\log_{10}{8} &\approx& \frac{12888}{14271}\approx 0.903\,090,\nonumber\\
\log_{10}{9} &\approx& \frac{13618}{14271}\approx 0.954\,243,\nonumber\\
\log_{10}{10} &=& \frac{14271}{14271} = 1, \nonumber \\
\label{log-ap}
\end{eqnarray}
In each case the rational approximation above gives the correct result
to within one part in two million and the correct decimal approximation
to all of the six digits given.  (The errors for the common logarithms
are smaller than for the natural logarithms because the ratios of the
natural logarithms are given by the procedure above more precisely than
is the procedure for approximating $\ln{2}$ and $\ln{10}$.)  Thus one
can easily use the formulas given in this paper to calculate by hand the
common logarithms of all integers up through ten to six significant
digits correctly.  (The most difficult part without using a calculator
is doing the divisions to convert the rational approximations to decimal
approximations, which I admit I haven't done.)

One could extend this method to get rational approximations for the
natural and common logarithms of the primes 11, 13, 17, 19, etc.\ by
using such facts as $2\ln{11} = \ln{(121/120)} + \ln{120} \approx 2/241
+ 82h - 6g$, $2\ln{13} = \ln{(169/168)} + \ln{168} \approx 2/337 +
\ln{7} + 55h + g$, $2\ln{17} = \ln{(289/288)} + \ln{288} \approx 2/577 +
98h + 2g$, $2\ln{19} = \ln{(361/360)} + \ln{360} \approx 2/721 + 102h -
5g$, etc., but these shall be left as exercises for the reader.

\newpage

\section{Conclusions}

The fact that the Kirnberger kernel or Kirn-kern, one-twelfth the atom
of Kirnberger, or the difference between a grad and a schisma, is very
small, may be attributed to the fact that one may calculate that the
ratio of (a) the difference between four equal-tempered semitones and a
justly tuned major third ($\ln{2^{1/3}}-\ln{(5/4)}$) and of (b) the
difference (the grad) between a justly tuned perfect fifth and seven
equal-tempered semitones ($\ln{(3/2)}-\ln{2^{7/12}}$), is very nearly
\begin{equation}
\frac{4(81+80)(128-125)}{3(81-80)(128+125)-(81+80)(128-125)}
=\frac{1932}{276},
\label{integer}
\end{equation}
which is an integer, 7.  Then when one writes the Kirn-kern as a natural
logarithm as $k = (\ln{2^{1/3}}-\ln{(5/4)}) -7(\ln{(3/2)}-\ln{2^{7/12}})
= (161/12)\ln{2} - 7\ln{3} - \ln{5} =
(1/12)[11\ln{(128/125)}-21\ln{(81/80)}] =
(1/6)[11\tanh^{-1}{(3/253)}-21\tanh^{-1}{(1/161)}]$ in terms of
power-series expansions of the inverse hyperbolic tangents in the small
quantities $x = (128-125)/(128+125) = 3/253$ and $y = (81-80)/(81+80) =
1/161$ for $\ln{(128/125)}$ and $\ln{(81/80)}$, one finds that because
$11(3/253)-21(1/161) = 0$, the first-order terms cancel, and one is left
with cubic and higher-order terms in $x$ and $y$, making $k$ very
small.  The cubic terms give $k \approx (1/18)(11x^3 - 21y^3) =
160/216\,414\,429 \approx 7.4\times 10^{-7}$.

One may use similar approximations to get rational approximations for
the natural logarithm of the semitone, $h = (1/12)\ln{2} \approx
716/12397$, and of the grad, $g = \ln{(3/2)}-\ln{2^{7/12}} = \ln{3} -
(19/12)\ln{2} \approx 14/12397 = 2/1771$.  From these, one may get
rational approximations to the natural and common logarithms of the
product of any integer powers of a semitone, of 3, and of 5.

\section*{Acknowledgments}

This paper was motivated by my attending a music class that my sons took
over a decade ago.  I learned virtually no music but became interested
in the ratios of the frequencies of notes.  Later I stumbled upon the
`coincidence' I finally have been able to explain here from the fact
that I read in two different dictionaries two different definitions of
the schisma (one actually for what is more standardly called the grad,
as I learned later) that numerically were very nearly equal.  I posted
the ratio to 1024 digits on my door and offered \$10 for anyone who
could discover the word whose two definitions I had found had that
ratio.  Eventually Andrzej Czarnecki solved the riddle and won the \$10
by entering the right number of digits of the ratio I had given into
Google to be led to the mystery word, schisma.  This research was
supported in part by the Natural Sciences and Engineering Research
Council of Canada.

\end{document}